# Subwavelength line imaging using plasmonic waveguides

Nina Podoliak[1], Peter Horak[1], Jord C. Prangsma[2], and Pepijn W. H. Pinkse[3]

1) Optoelectronics Research Centre, University of Southampton, Southampton, SO17 1BJ, UK. 2) Nanobiophysics (NBP), MESA+ Institute for Nanotechnology, University of Twente, P.O. Box 217, 7500 AE Enschede, The Netherlands. 3) Complex Photonic Systems (COPS), MESA+ Institute for Nanotechnology, University of Twente, P.O. Box 217, 7500 AE Enschede, The Netherlands.

**We investigate the subwavelength imaging capacity of a two-dimensional fanned-out plasmonic waveguide array, formed by air channels surrounded by gold metal layers for operation at near-infrared wavelengths, via finite element simulations. High resolution is achieved on one side of the device by tapering down the channel width while simultaneously maintaining propagation losses of a few dB. On the other, low-resolution side, output couplers are designed to optimize coupling to free space and to minimize channel cross talk via surface plasmons. Point sources separated by $\lambda/15$ can still be clearly distinguished. Moreover, up two 90% of the power of a point dipole is coupled to the device. Applications are high-resolution linear detector arrays and, by operating the device in reverse, high-resolution optical writing.**

*Index Terms*— High-resolution imaging, plasmon waveguides, surface plasmons.

## I. INTRODUCTION

Linear detector arrays are popular in many applications ranging from spectrometers and particle counters to position encoders and autofocussing systems. The achievable resolution can be limited by a variety of factors, such as the detector pixel size or the magnification of the imaging system, but ultimately for free-space optical methods it is determined by the diffraction limit of light. Vice versa, optical encoding of linear systems like long nanoparticles [1–4] and the detection of fluorescent markers along linear biological systems such as DNA has the same resolution limitations [5]. The diffraction-limited resolution can be improved by using high-index liquids for immersion microscopy, solid immersion lenses [6], high-index scattering media [7], and by superresolution techniques such as STED, STORM or PALM [8].

An alternative method for achieving high resolution makes use of plasmons in metals. The most recent developments exploit metamaterials with a negative refractive index to form "superlenses" and "hyperlenses" [9]. A more classic method exploits an array of subwavelength metal wires operating in "canalization" regime [10–13]. In this method plasmons propagating along metallic wire arrays can carry and transform a subwavelength scaled field and thus can be used to get high resolution images e.g., in near-field microscopy. The price to pay for systems with metals is either high losses due to the below-cut-off operation of subwavelength waveguides or unwanted cross coupling between closely-spaced nanowires for metal guiding, an effect already studied for 3D plasmonic wire arrays [13]. Another limitation to such a metallic wire array is that the length of the wires should obey the Fabry-Pérot resonance condition, which also introduces limitation to the operation bandwidth [13]. We show here that the 2D geometry profits from the best of two worlds: By making a stack of thin parallel dielectric waveguides in a metal, the resolution along one dimension can be subwavelength by the confining effect of the metal, while simultaneously the losses for plasmons polarized perpendicular to these 'ribbons' are small.

In this paper we design and optimize in detail a subwavelength plasmonic line image transformer for near-field imaging. The device consists of a two-dimensional fanned-out plasmonic waveguide array that transmits and magnifies a near field on a pixel-by-pixel basis creating a discrete image. Because of this we avoid calling the devise a "lens" (or "hyperlens"), and use the term "image transformer" or "imaging system" instead. We investigate propagation losses and thus corresponding maximum device lengths depending on waveguide dimensions and study how short, tapered sections at the input can increase resolution while still maintaining low overall losses and small channel cross coupling. At the device output we introduce a properly tailored output coupler that enhances optical coupling to free space. This is shown to be an effective method to reduce back-reflection, thus minimizing Fabry-Pérot effects, and also to suppress channel to channel coupling at the output facet via surface plasmon waves. Finally we show that two point dipole sources separated by $\lambda/15$ can be effectively distinguished by such a device.

## II. DESIGN OF SUBWAVELENGTH PLASMONIC WAVEGUIDES

Utilizing plasmonic waveguides is an established method for confining light to a subwavelength scale. There are two principal types of plasmonic waveguides known as metal slab waveguides, consisting of a thin metal stripe within a dielectric or air cladding, and metal-clad waveguides, consisting of a guiding dielectric channel surrounded by metal [14–16]. Although the metal slab waveguide was shown to support a low attenuation propagating mode in the form of long-range surface plasmon polaritons, low attenuation is achieved at the cost of reduced confinement [14,15]. Thus, to avoid cross coupling between two parallel waveguides of this type they must be separated by a relatively large distance, of the order of one wavelength. On the other hand, in metal-clad waveguides, the light is mainly concentrated in the dielectric channel and extends only a few tens of nanometers into the metal allowing micron-scale propagation with nanometer-scale confinement [15,16]. For example, the plasmon penetration depth into gold on a gold-air interface at $\lambda$=1550 nm can be estimated as $\delta = \frac{\lambda}{2\pi\sqrt{-\varepsilon'_m}} \approx 21.5$ nm [17], where the dielectric constant of gold at this wavelength is $\varepsilon'_m = -132$ [18]. Therefore a relatively thin subwavelength layer of gold (around 50 nm) will be sufficient to optically isolate two air waveguides. For this reason the metal clad type of waveguides is our choice for constructing an imaging device with subwavelength image resolution.

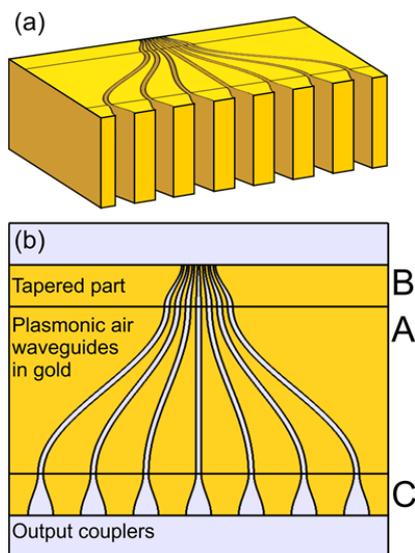

Fig. 1. Schematic geometry of plasmonic line imaging system (a) in perspective and (b) in top view. In the center of the device (part A) thin air channels are fanning out to magnify an input image. On the high-resolution side the waveguides are tapered down to achieve even smaller spacing (part B), whereas on the low-resolution side the waveguides are terminated by funnel-like output couplers (part C).

Fig. 1 shows a schematic of the plasmonic line imaging system. One could consider it as a subwavelength plasmonic version of a pitch conversion device [19]. A similar planar fanned array of waveguides was proposed for a dark field microscope [20]. Our device consists of a row of tailored air guiding plasmonic waveguides in a gold slab. In the main part of the device (part A in Fig. 1), the waveguides have constant widths but their separation increases gradually from the input towards the output. Curved, S-shaped waveguides are used to ensure that at input and output all waveguides are orthogonal to the device edge and thus exhibit the same input and output coupling efficiencies. At the input (part B), the waveguide widths are tapered down to achieve closer packing and thus higher resolution. At the output (part C), the waveguides are terminated by a set of funnel-like couplers allowing effective field radiation into free space. Below we will discuss the optimization of parts A, B and C of the imaging system to minimize losses and to ensure that there is no cross coupling between the waveguides. By suppressing cross coupling between the waveguides we ensure that each waveguide transmits optical fields independently and acts as an individual "pixel". The device thus effectively maps a near field at the high-resolution input to the low-resolution output, allowing in this way image magnification. However, the device can also work in the opposite way, mapping the optical field of a large object onto the subwavelength scale. Note that as an image is transferred by propagating plasmons, the device operates at TM incident polarization, i.e., at polarization in the image plane of Fig. 1(b).

In principle, broadband operation in both infrared and visible light is possible as the operation principle does not rely on a resonance. Gold is known to be a good metal for operation in the near infrared, while silver exhibits lower absorption in the visible part of the spectrum. In the following we focus on device design for operation in the near-infrared, specifically at a wavelength of 1550 nm.

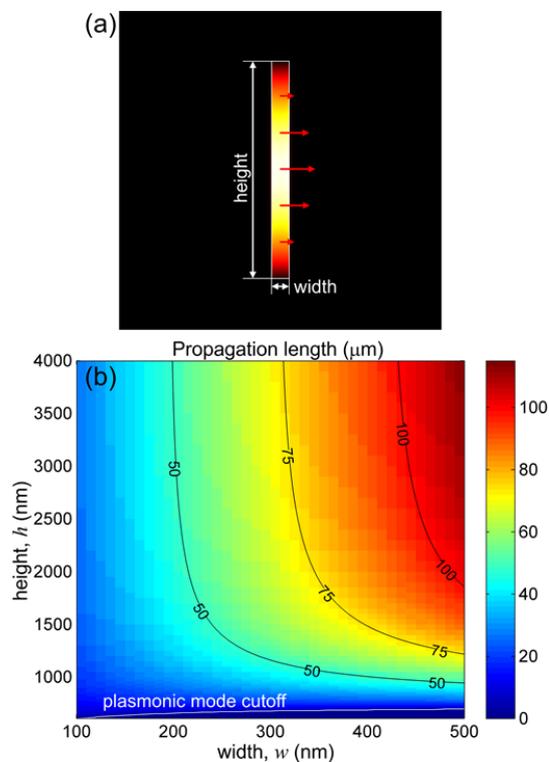

Fig. 2. (a) Normalized electric field profile and polarization (indicated by the red arrows) of the fundamental plasmonic mode in a rectangular aperture. (b) Propagation length (distance at which optical field magnitude decreases by a factor of $e$) of the fundamental mode supported by a rectangular air hole in gold depending on the waveguide dimensions at 1550 nm wavelength.

Plasmon propagation is usually associated with large losses compared to e.g. dielectric waveguides. We thus start our discussion by analyzing optical losses of the fundamental plasmonic mode supported by a subwavelength rectangular air aperture in gold [21] depending on the aperture height and width, shown in Fig. 2(a). Numerical simulations were performed using a fully vectorial finite-element method (Comsol Multiphysics) for a working wavelength of 1550 nm assuming a gold refractive index of 0.55+11.5$i$ [18]. The calculated mode profile and propagation length, defined as the distance at which the optical field amplitude decreases by a factor of $e$, depending on the waveguide dimensions, are shown in Fig. 2. For a given targeted propagation length (i.e. a contour line in Fig. 2(b)) a range of waveguide sizes can be chosen. Choosing a waveguide with a high aspect ratio (ratio between height and width) allows the transverse dimension to be of subwavelength values. In other words, the cost of increasing resolution in one direction is a reduction of resolution in the other direction. For example, a propagation length of 50 μm can be achieved by waveguide cross sections of 400 nm × 1 μm or 200 nm × 3 μm. If we target a linear imaging array with high resolution in one direction and low transmission losses, the suitable waveguide geometry for the waveguides in part A (Fig. 1) of the device is thus a thin air slit of thickness 200-300 nm and height >3 μm.

Such a choice of waveguide widths and with individual waveguides separated by sufficiently thick metal layers to avoid optical cross coupling will limit the periodicity of the structure at the input of part A to about 400 nm. A further increase in resolution can only be achieved by thinner waveguides albeit at the cost of increased losses, as shown by Fig. 2. However, this may be acceptable for the device if the length of such narrow channels is kept as short as possible. We therefore consider a short section of the device at the input (part B in Fig. 1) where the waveguide thickness is tapered down. Specifically, we modelled light propagation in the central channel of a 2 μm long tapered part B, the geometry of which is sketched in the inset of Fig. 3(a) and estimated losses in the channel depending on the taper dimensions at the input (air channel width $d_0$, and gold layer thickness $g_0$). The channel width at the output of the taper (boundary between parts B and A) was fixed to $d$ = 250 nm and the gold layer thickness between channels was $g$ = 250 nm (total periodicity of 500 nm). Note that the channel width could be significantly reduced by using a high index dielectric instead of air.

The high aspect ratio of the waveguide in this geometry allowed us to reduce the complexity of the problem and perform rigorous simulations in only two dimensions; the calculated losses induced by the tapered part B and the fraction of optical power cross-coupled between channels are shown in Fig. 3. Losses depend on the air channel width at the input but are almost independent of the gold layer thickness in-between channels for thicknesses above 75 nm. However, by reducing the gold layer thickness, cross talk between channels increases, as shown in Fig. 3(b) by the fraction of input power cross-coupled to the other channels. When the gold layer thickness is 50 nm or more, less than 0.4% of power is coupled between channels. 2 μm long sections of a waveguide with a constant width of $d_0$ = 250 nm exhibit a loss of 0.27 dB. If the waveguide thickness is tapered down from 250 nm to 50 nm at the input, ~0.3 dB of additional loss appears. Note that overall propagation losses of 0.55 dB as shown in Fig. 3 correspond to huge losses of ~2750 dB/cm, but over the short distance of 2 μm these are nearly negligible. By tapering the waveguide thickness down to 50 nm and having a 50 nm separation between the waveguides, a periodicity of 100 nm (≈$\lambda$/15) can thus be achieved at the input of the device (input of part B) while still maintaining moderate overall propagation losses below 1 dB and negligible cross coupling.

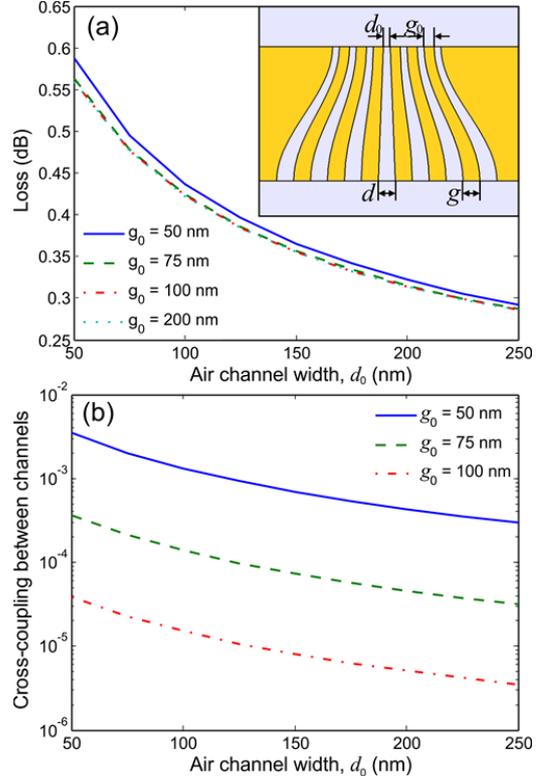

Fig. 3. (a) Losses and (b) the fraction of input power cross-coupled between channels induced by the tapered part B of the device depending on air channel width at the top $d_0$ and gold layer thickness between two channels $g_0$. Air channel width at the output of the tapered part is 250 nm, and the gold layer thickness is 250 nm (periodicity of 500 nm). Inset shows geometry of part B.

Next, we consider the design of the output port of the device, part C in Fig. 1. Fig. 4(a) shows the electric field amplitude of the propagating wave inside a straight and 250 nm wide waveguide excited by a point dipole source at the input (top of the figure) at wavelength $\lambda$=1550 nm. Two problems appear with abrupt terminating the plasmonic waveguide at the output: (i) Back reflection of plasmons at the gold-air interface at the output (bottom) creates a standing wave inside the waveguide and thus makes transmittance dependent on waveguide length and on wavelength. (ii) A close inspection of the electric field at the output (not visible in Fig. 4(a)) shows the excitation of propagating plasmons at the bottom surface of the device that will cause cross coupling between waveguides and thus will reduce the signal-to-noise ratio of the transmitted image.

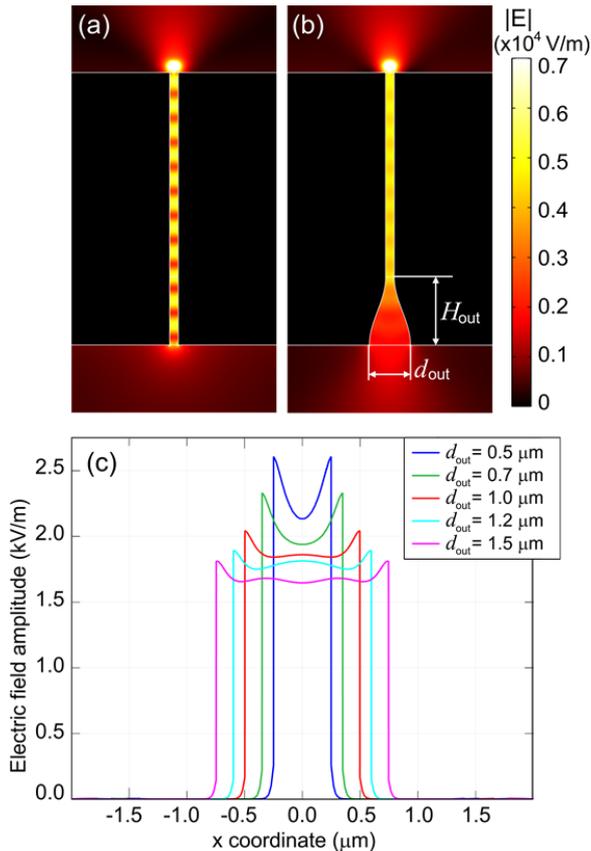

Fig. 4. Amplitude of the electric field inside a straight 250 nm wide waveguide (a) without and (b) with output coupler. Significant reduction of the standing-wave amplitude is observed with the coupler. Coupler size $H_{out}$ = 2 μm, $d_{out}$ = 1.2 μm. (c) Electric field profile at the coupler output for different output widths $d_{out}$ and for $H_{out}$ = 2 μm.

Both of these problems can be significantly reduced by introducing an output coupler reminiscent to a microwave horn antenna at the end of each channel. We modelled light propagation in a straight, 250 nm wide waveguide with a coupler at the output, see Fig. 4(b). The coupler shape was described by two sine-shaped lines allowing for a smooth design without sharp corners. The symmetric shape of the coupler ensures that two plasmons propagating along the opposite surfaces remain in phase. The coupler is characterized by two parameters, its length $H_{out}$ and its output width $d_{out}$, which we can optimize for minimum back reflection. As the waveguide widens, standard dielectric TM slab waveguide modes come into existence as they cross cutoff, and light propagating in the plasmonic modes of the narrow waveguide can be transferred adiabatically into the symmetric waveguide modes. The first-order mode was observed at the coupler output when its width was around 1 μm, while the formation of a higher-order mode was seen when the output width was above 1.5 μm, see Fig. 4(c). The electric field profile of the first-order mode has a maximum in the middle of the channel and decays at the edges. As most of the light in the waveguide mode propagates in the middle of the channel, it would not "see" the gold-air interface, and thus the back reflection is reduced. A minimum of standing-wave amplitude along the waveguide in Fig. 4(b) was found at $H_{out}$ = 2 μm. For low back reflection and mostly fundamental-mode output the optimal coupler output width of $d_{out}$ = 1.2 μm is chosen in the following. By using an output coupler of these optimal dimensions, the amplitude of the standing wave in the narrow part of the channel was reduced by a factor of 6.5. The residual back reflection seen in Fig. 4(b) is caused by some fraction of light still propagating in the plasmonic modes along the output coupler edges. Significant reduction of the surface waves at the bottom edge of the device is also observed with the use of the output coupler.

An alternative solution for reducing back reflection, i.e. optimizing transmission, could utilize plasmonic antennas at the output. Suitable antenna geometries that improve light radiation into free space have already been reported [22, 23]. It was shown that a significant reduction in back reflection can be achieved when resonance and impedance-matching conditions between antennas and waveguides are met. By improving light radiation into free space, antennas would also reduce light coupling to the surface waves along the bottom surface. However, employing antennas, whose performances are based on a resonance, can introduce limitations to the operation bandwidth of the device and their design will in general be more challenging to fabricate than the simpler output funnel proposed in Fig. 4. A further option for the output coupling would be to employ Si–plasmonic couplers [24] to couple the optical field from the plasmonic waveguides to Si waveguides which then transmit the light to the detectors [20].

## III. Device Imaging Operation and Resolution

To illustrate the imaging properties of the fanned-out plasmonic waveguide array, we numerically simulated imaging two electric dipole point sources radiating at 1550 nm wavelength separated by a subwavelength distance. The dimensions of the plasmonic waveguides are the following. At the high-resolution input side (part B in Fig. 1) the air channels are 100 nm thick with a 150 nm gold layer between them (i.e., a periodicity of 250 nm). The tapered part is 2 μm long, along which the waveguides are widened to 250 nm and the gold layer thickness becomes 250 nm. In the main part of the device (part A) the fanned-out waveguides of a constant thickness are bent to increase the channel separation to 2.5 μm (total magnification factor of 10). To simplify the geometry, we modelled 7 waveguides forming a symmetrical structure and assumed the central straight waveguide to be 8 μm long. There is an output coupler (part C of the device) at the end of each waveguide (2 μm length, 1.2 μm output width).

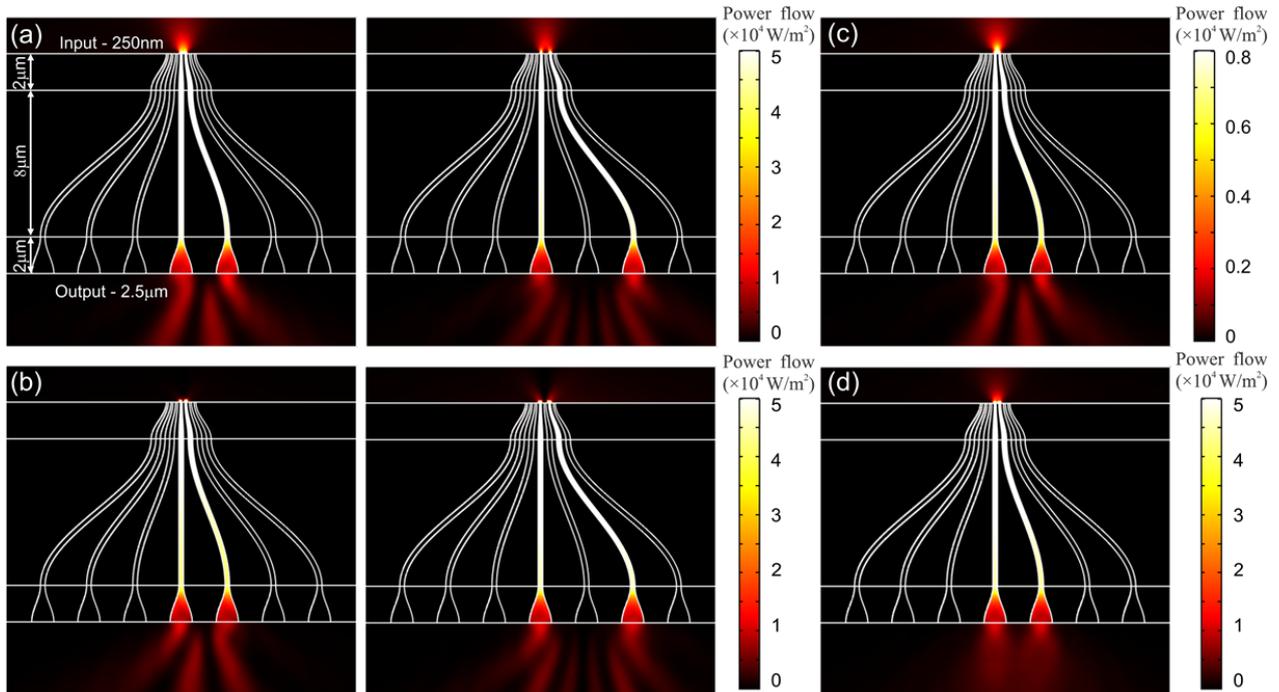

Fig. 5. Power flow inside the plasmonic imaging system in the case of (a) two coherent point dipole sources of equal amplitude and phase separated by 250 nm (left) and 500 nm (right), (b) two sources of equal amplitude and opposite phase separated by 250 nm (left) and 500 nm (right), (c) a single source positioned in-between two channels, (d) two incoherent point dipole sources separated by 250 nm. In all cases the sources are positioned at an optimal distance of 90 nm from the top surface.

Firstly, we find an optimal distance of a single point source from the input interface at which coupling of the light into the corresponding channel is maximized. Objects in the plane at this distance in front of the waveguide array can be imaged with the best possible resolution: when the source is placed behind this plane, the light spreads out before reaching the waveguide facet and thus couples to more than one waveguide; likewise, a source too close to the input excites plasmonic surface waves at the front of the device which also couple light to other channels, decreasing the image contrast at the output. We found that the optimal distance depends on the waveguide width and periodicity at the input: smaller periodicity requires an object to be placed closer to the device. For example, for the parameters chosen here (100 nm thick waveguides with 150 nm gold layer separation), the optimal object distance is 90 nm. When the waveguide width is reduced to 50 nm and the gold layer thickness to 50 nm, the optimal distance becomes 30 nm.

Secondly, we show that optical fields from an emitter can be efficiently coupled into subwavelength plasmonic waveguides. Previous studies showed that due to the tight confinement the optical emission can be almost entirely coupled to the propagating plasmon modes of metallic nanowires [25]; this strong enhancement of the emission from a source in the vicinity of the plasmonic wire is due to the Purcell effect. Strong coupling between quantum dots and metallic nanowires was also demonstrated experimentally [26]. Here, we calculated the coupling efficiency between a point source with a fixed dipole moment and the plasmonic waveguide of the modelled geometry. A coupling efficiency of almost 80% was obtained for a point source positioned at 90 nm distance from waveguides with 250 nm periodicity at the input, and over 90% efficiency for a source positioned at 30 nm distance from waveguides with 100 nm periodicity.

Fig. 5 shows the power flow of propagating waves inside the plasmonic imaging waveguides excited by one or two point electric dipole sources. Two point sources of equal amplitude are placed at the optimal distance of 90 nm from the input, centered in front of channels and separated by 250 nm and 500 nm, corresponding to one and two waveguide array periods, respectively. Images created by two coherent in-phase sources are shown in Fig. 5(a), while Fig. 5(b) presents images of two out-of-phase sources. By standard optical microscopy, two point sources separated by these subwavelength distances are indistinguishable: the in-phase sources appear as a single one, while radiation from the out-of-phase sources interferes destructively and nearly perfectly cancels in the far field, as can be seen at the top part of the figures. However, placed at the optimal distance from the waveguide input, the light from each source in all cases couples effectively and independently into a single channel and propagates towards the output where it is re-emitted into free space, creating interference patterns. At the output side of the imaging system the two sources are separated by well over one wavelength and can thus be clearly resolved.

We also performed calculations for a single source placed in-between two channels, see Fig. 5(c). In this case the light is coupled to two neighboring waveguides, generating output equivalent to the case of two coherent sources positioned in front of two neighboring channels. Thus, two coherent sources can be clearly distinguished from a single source only if they are separated by another unilluminated channel in-between. In other words, the system resolution is given by twice the waveguide periodicity at the input. In the current geometry two coherent point sources separated by 500 nm (<$\lambda$/3) can be unambiguously distinguished by the optical system. With further reducing the input taper periodicity down to 100 nm (see Fig. 3), 200 nm (<$\lambda$/7.5) resolution can be achieved. However, if the sources are non-coherent, see Fig. 5(d), the case of two sources separated by the waveguide periodicity can be clearly distinguished from the case of one source in-between channels by the absence of an interference pattern at the output (as the waveguide separation at the output is larger than the wavelength). Thus, twice smaller resolution is achieved in the general case of non-coherent sources, namely $\lambda$/15 for the 100 nm input periodicity.

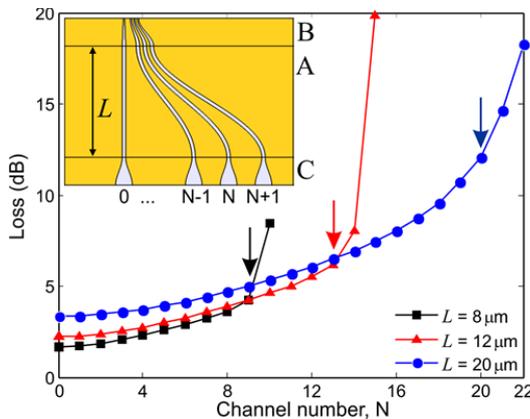

Fig. 6. Propagation loss of each channel depending on the channel number calculated for three lengths, $L$. Arrows mark points where the cross talk between channels occurs due to sharp bending. Inset shows the geometry.

Finally, we calculated total propagation losses of the device. Fig. 6 shows losses of each channel depending on the channel number calculated for three device lengths $L = 8$, 12, and 20 µm. We also analyzed losses induced by each part of the channel, parts A, B, and C, separately. As expected, the loss of the output coupler (part C) is the same in each channel and equals to 0.14 dB, while losses of parts A and B gradually increase in channels positioned further away from the center because of their longer lengths. The loss induced by part A also increases proportional to the devise length $L$. For example, the total loss of the central channel is 1.7 dB and increases up to 4.2 dB in the 9[th] channel at $L = 8$ µm; while losses of 2.24 dB and 6.2 dB are induced by the central and 13[th] channel, respectively, at $L = 12$ µm.

The rapid increase in losses (marked by arrows in Fig. 6) is associated with light leaking to neighboring channels due to sharp channel bending in part A. The channel thickness in part A is constant, so the tilt and bend lead to narrowing of the gold layer between two channels, which at some point becomes too thin and cross coupling between channels occurs. This effect therefore limits the total number of channels in the device. By increasing the device length, the allowed number of channels can be increased. For example, at $L = 8$ µm the device can contain up to 19 channels, while at $L = 12$ µm the total number of channels can be 27. At $L = 20$ µm, however, the total channel number is restricted to 41 by the rapid increase of losses in part B rather than in part A. The total loss of channel 20 reaches 12 dB. For a further increase in channel number, a wider channel width would have to be chosen to decrease losses, see Fig. 2(b).

Finally, we tested the spectral range of the device of Fig. 5. A priori we expect the device to have a broad operation range, since it does not depend on resonances and since the design is optimized to suppress reflections and interference. Indeed we found consistent operation in the whole wavelength range from approximately 1 µm to above 2 µm. At wavelengths below 1 µm the device performance is limited by gold material absorption and thus a device operating in the visible would have to be based on a different material choice. At wavelengths above 2 µm cross-coupling between channels occurs. However, the device dimensions could be easily optimized for operation in the mid-infrared part of the spectrum.

## IV. CONCLUSION

In conclusion, we have demonstrated by finite-element simulations that it is possible to make a linear array of air guided plasmonic waveguides with fanned out geometry to create a magnifying line image transformer device. The device can effectively couple a large fraction (up to 90%) of the light from small objects or emitters to tapered plasmonic waveguides on a high-resolution side. The waveguides transmit the signal to a low-resolution side via propagating plasmon modes without cross talk and with moderate losses, magnifying the image. The output couplers are designed to enhance coupling to free space. A resolution of $\lambda$/15 is achieved. Experimental realization of the structures investigated here is possible with existing 3D lithography systems [27]. Alternatively, deep reactive ion etching could be used to cut the tapered channels in a few micrometers deep gold layer. The device can be applied as a high-resolution linear detector or, by operating the in reverse, high-resolution optical writing.

This work was supported by the U.K. Engineering and Physical Sciences Research Council EPSRC (grant no. EP/J012874/1), Stichting Technische Wetenschappen (STW) under the nanoscopy program (project nr.12149), and the Nederlandse Organisatie voor Wetenschappelijk Onderzoek (NWO) via a VICI grant.